\documentclass[%
reprint,
 amsmath,amssymb,
 aps,
prl,
]{revtex4-1}

\usepackage{graphicx}
\usepackage{array}
\usepackage{dcolumn}
\usepackage{bm}
\usepackage{listings}
\usepackage{scalefnt}
\usepackage{xcolor}
\usepackage{booktabs}
\usepackage{multirow}
\usepackage{tensor}
\usepackage{hyperref}
\usepackage{mathtools}
\hypersetup{
    colorlinks,
    linkcolor={blue!80!black},
    citecolor={blue!80!black},
    urlcolor={blue!80!black}
}
\usepackage{color}

\usepackage{physics}
\usepackage[pagewise]{lineno}
\usepackage{nicefrac}

\def\noteC#1{\textbf{\color{green}}} 

\begin{document}
\title{Heisenberg-limited metrology via weak-value amplification\\ without using entangled resources}
\author{Yosep Kim}
\thanks{yosep9201@gmail.com}  
\affiliation{Department of Physics, Pohang University of Science and Technology (POSTECH), Pohang 37673, Korea}
 
\author{Seung-Yeun Yoo}
\affiliation{Department of Physics, Pohang University of Science and Technology (POSTECH), Pohang 37673, Korea}
 
\author{Yoon-Ho Kim}
\thanks{yoonho72@gmail.com}  
\affiliation{Department of Physics, Pohang University of Science and Technology (POSTECH), Pohang 37673, Korea}

\date{\today}
             
\begin{abstract}
	Weak-value amplification (WVA) provides a way for amplified detection of a tiny physical signal at the expense of a lower detection probability. 
    	Despite this trade-off, due to its robustness against certain types of noise, WVA has advantages over conventional measurements in precision metrology. Moreover, it has been shown that WVA-based metrology can reach the Heisenberg-limit using entangled resources, but preparing macroscopic entangled resources remains  challenging.
	Here we demonstrate a novel WVA scheme based on iterative interactions, achieving the Heisenberg-limited precision scaling without resorting to entanglement. 
	This indicates that the perceived advantages of the entanglement-assisted WVA are in fact due to iterative interactions between each particle of an entangled system and a meter, rather than coming from the entanglement itself.
	Our work opens a practical pathway for achieving the Heisenberg-limited WVA without using fragile and experimentally-demanding entangled resources.
\end{abstract}
\maketitle

Precise measurement of an interaction parameter is essential for studying various quantum physical effects. The interaction parameter can be estimated from the state change of a meter after interacting with a system. Since the system and the meter are generally entangled by the interaction, post-selection of the system collapses the meter's state: an elaborate post-selection scheme enables amplified detection of the interaction parameter. 
	In the weak interaction regime, the amplification factor corresponds to the so-called weak value, and thus this metrological protocol is known as weak-value amplification (WVA)~\cite{Aharonov88,Dressel14,Cho10}. WVA has been actively deployed in precision metrology to amplify tiny physical effects, such as the optical spin Hall effect \cite{Hosten08}, an ultra-small phase shift \cite{Strubi13,Xu13},  and the nonlinear effect due to a weak beam of light \cite{Feizpour11,Hallaji17,Chen18}. 

WVA provides enhanced sensitivity over conventional measurements (in the sense that small signals are physically amplified), but the enhancement does not necessarily result in a higher signal-to-noise ratio (SNR) due to the reduced detection probability by post-selection, which in turn causes a greater statistical error \cite{Combes14,Ferrie14,Knee14,Zhang15}. 
	Despite this trade-off, WVA-based metrology has attracted a lot of attention as it offers meaningful robustness against certain types of noise, e.g., temporally correlated noises~\cite{Feizpour11,Sinclair17}, detector jitter~\cite{Kedem12,Jordan14,Viza15}, and detector saturation \cite{Harris17, Xu20}. 
	 In addition, there are proposals to achieve the Heisenberg-limited precision scaling in WVA-based metrology using quantum resources such as squeezing~\cite{Pang15S} and entanglement~\cite{Pang14,Pang15E}. Recently, the entanglement-assisted WVA has been demonstrated in photonic systems~\cite{Chen19,Starek20}, but the practical metrological applications are limited by the difficulty in preparing macroscopic entangled resources.
	 	 
Here we demonstrate a novel WVA scheme based on iterative interactions, achieving the Heisenberg-limit without resorting to entanglement. $N$ iterative interactions enhance the post-selection probability by $N^2$ while keeping the amplification factor unchanged. In a proof-of-principle experiment, we estimate a path-dependent polarization change using our WVA scheme, showing a good agreement with the Heisenberg-limited precision scaling. 
 
\begin{figure}[t]
\centering
\includegraphics[width=0.9\columnwidth]{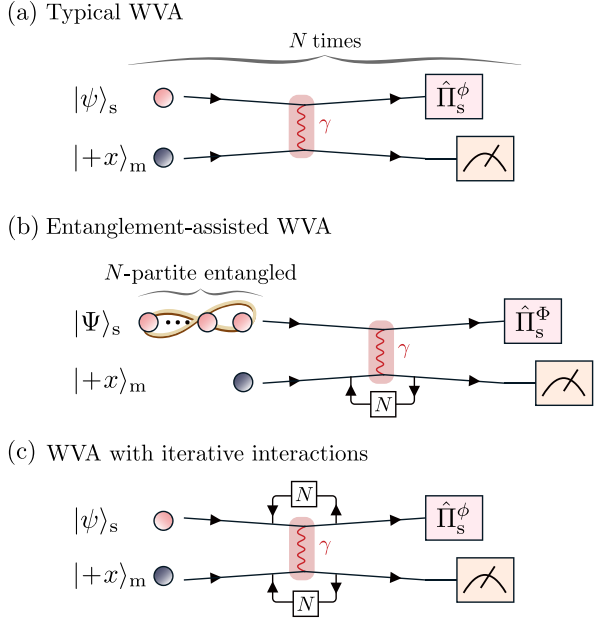}
\caption{Three weak-value amplification (WVA) scenarios for estimating an interaction strength $\gamma$ with a total of $N$ interactions between a system and a meter. (a) The meter state is measured independently $N$ times after a single interaction with a system particle and the post-selection of the system. (b) The meter state is measured once after $N$ consecutive interactions with each particle of an $N$-partite entangled system and the post-selection of the system.
(c) The meter state is measured once after $N$ iterative interactions with a system particle and the post-selection of the system.}
\label{fig1} 
\end{figure}

\textit{Theory}.---We first briefly review a typical WVA-based metrology depicted in Fig.~\ref{fig1}(a). Given the initial system state as $|\psi\rangle_\text{s}$ and the initial meter state as $|\!+\!x\rangle_\text{m}$, the weak interaction of 
$\hat{U}=e^{i\gamma\hat{A}\otimes\hat{\sigma}_z}$ changes the system-meter state to,
\begin{equation}
\hat{U}|\psi\rangle_\text{s}|\!+\!x\rangle_\text{m} \approx |\psi\rangle_\text{s}|\!+\!x\rangle_\text{m}+i \gamma \hat{A}|\psi\rangle_\text{s}|\!-\!x\rangle_\text{m},
\label{eq2}
\end{equation}
where $\hat{A}$ is the system observable, $\hat{\sigma}_z$ is the Pauli-$z$, and $|\!\pm\!x\rangle_\text{m}$ are the eigenstates of the Pauli-$x$. Assuming the initial states and the system observable are known, the interaction parameter $\gamma$ is estimated from the change of the meter state. Moreover, the state change can be significantly amplified by projecting the system onto $|\phi\rangle_\text{s}$ and post-selecting the outcome, 
\begin{equation}
_\text{s}\langle\phi|\hat{U}|\psi\rangle_\text{s}|\!+\!x\rangle_\text{m}\approx\ \!_\text{s}\langle\phi|\psi\rangle_\text{s}(|\!+\!x\rangle_\text{m}+i\gamma\langle\hat{A}\rangle_\text{w}|\!-\!x\rangle_\text{m}).
\label{eq3}
\end{equation}
The post-selection probability is $P\approx|{}_\text{s}\langle\phi|\psi\rangle_\text{s}|^2$, and the amplification factor corresponds to the weak value~\cite{Aharonov88},
\begin{equation}
\langle\hat{A}\rangle_\text{w}=\frac{_\text{s}\langle\phi|\hat{A}|\psi\rangle_\text{s}}{_\text{s}\langle\phi|\psi\rangle_\text{s}}.
\label{WV}
\end{equation}
The weak value can be outside the eigenvalue spectrum of $\hat{A}$ and is generically a complex number. These intriguing features have been extensively studied in the context of amplified detection~\cite{Hosten08,Strubi13,Xu13,Feizpour11,Hallaji17,Chen18} and, more recently, in the context of quantum information science, e.g., quantum tomography \cite{Lundeen11,Kim18}, quantum foundation \cite{Kocsis11,Mahler16,Cho19,Ramos20,Kim21}, and decoherence management \cite{Kim12}.


When repeating $N$ independent WVA measurements of the meter state in Eq.~(\ref{eq3}), the minimum achievable uncertainty of an~unbiased estimator $\gamma$ is given by the quantum Cram\'{e}r-Rao bound~\cite{Braunstein94,Braunstein96},
 \begin{equation}
\Delta\gamma_\text{ind}^{(N)} \geq \frac{1}{\sqrt{NPI(\gamma)}},
\label{eq5}
\end{equation}
where $P$ is the post-selection probability and $I(\gamma) \approx 4|\langle\hat{A}\rangle_\text{w}|^2$ is the quantum Fisher information contained in the meter state under the weak interaction condition of $|\gamma| \ll |1/\langle\hat{A}\rangle_\text{w}|$, see Supplemental Material for the details. The result in Eq.~(\ref{eq5}) shows that the measurement uncertainty of a typical WVA-based metrology scales with $1/\sqrt{N}$ according to the standard quantum limit \cite{Giovannetti04}. To achieve a large amplification factor in Eq.~(\ref{WV}) and large quantum Fisher information, the post-selection state $|\phi\rangle_\text{s}$ has to be set to reduce a post-selection probability $P\approx|{}_\text{s}\langle\phi|\psi\rangle_\text{s}|^2$. Considering the uncertainty bound in Eq.~(\ref{eq5}), the perceived enhancement from the amplified detection is, therefore, inevitably compromised by the small detection probability. However, note again that the amplified detection enables us to suppress certain types of noise~\cite{Feizpour11,Sinclair17,Kedem12,Jordan14,Viza15,Harris17, Xu20}. 


We can leverage an entangled system to overcome the standard quantum limit in Eq.~(\ref{eq5})~\cite{Pang14,Pang15E,Chen19,Starek20}. The entanglement-assisted WVA scheme is depicted in Fig.~\ref{fig1}(b). Each particle of the $N$-partite entangled system in $|\Psi\rangle_\text{s}$ interacts consecutively with the meter in $|\!+\!x\rangle_\text{m}$, and then the system is post-selected by $|\Phi\rangle_\text{s}$  for amplified detection of the interaction parameter $\gamma$. The interaction is written as $\hat{U}^{\otimes N}=e^{i\gamma\sum_{k=1}^N\hat{A}_k\otimes\hat{\sigma}_z}$, where $\hat{A}_k=\hat{\mathbb{I}}\otimes\dots\otimes\hat{A}\otimes\dots\otimes\hat{\mathbb{I}}$ denotes the observable $\hat{A}$ for the $k$-th system particle. According to the original proposal~\cite{Pang14}, the entangled system is initially prepared as a maximally entangled state and is post-selected as follows: 
\begin{eqnarray}
\label{ent_init}
|\Psi\rangle_\text{s}&=&\frac{1}{\sqrt{2}}(|\lambda_{\text{max}}\rangle_\text{s}^{\otimes N}+|\lambda_{\text{min}}\rangle_\text{s}^{\otimes N}), \\
|\Phi\rangle_\text{s}&\approx&\frac{1}{\sqrt{2}|\langle\hat{A}\rangle_\text{w}|}\bigg[(\langle\hat{A}\rangle_\text{w}^*-N\lambda_{\text{min}})|\lambda_\text{max}\rangle_\text{s}^{\otimes N}\nonumber\\
&&\ \ \ \ \ \ \ \ \ \ \ \ \ \ \ \ -(\langle\hat{A}\rangle_\text{w}^*-N\lambda_\text{max})|\lambda_\text{min}\rangle_\text{s}^{\otimes N}\bigg].\nonumber
\end{eqnarray}
Here $|\lambda_{\text{max}(\text{min})}\rangle_\text{s}$ is the eigenstate of $\hat{A}$ and has the maximum (minimum) eigenvalue of $\lambda_{\text{max}(\text{min})}$. Then, under the assumption of a large amplification factor and a small interaction strength,
\begin{equation}
|N\lambda_\text{max(min)}|\ll|\langle\hat{A}\rangle_\text{w}|\ll|\gamma|^{-1},
\label{condition}
\end{equation}
we can estimate the interaction parameter $\gamma$ with the enhanced post-selection probability, 
\begin{equation}
P_\text{ent}^{(N)}\approx |{}_\text{s}\langle\Phi|\Psi\rangle_\text{s}|^2\approx \frac{N^2(\lambda_\text{max}-\lambda_\text{min})^2}{4|\langle\hat{A}\rangle_\text{w}|^2},
\label{ent_prob}
\end{equation} 
while keeping the amplification factor in the meter state the same with Eq.~(\ref{eq3}), 
\begin{equation}
_\text{s}\langle\Phi|\hat{U}^{\otimes N}|\Psi\rangle_\text{s}|\!+\!x\rangle_\text{m}\approx
\ \!_\text{s}\langle\Phi|\Psi\rangle_\text{s}(|\!+\!x\rangle_\text{m}+i\gamma\langle\hat{A}\rangle_\text{w}|\!-\!x\rangle_\text{m}),
\label{ent_amp}
\end{equation}

For a fair comparison, the number of interactions in the entanglement-assisted WVA in Fig.~\ref{fig1}(b) needs to be the same with $N$ independent measurements of the typical WVA in Fig.~\ref{fig1}(a). To do so, we estimate the quantum Cram\'{e}r-Rao bound for a single entanglement-assisted WVA measurement to be,
\begin{equation}
\Delta\gamma^{(N)}_\text{ent}\geq \frac{1}{\sqrt{P^{(N)}_\text{ent}I(\gamma)}}  = \frac{1}{N|\lambda_\text{max}-\lambda_\text{min}|},
\label{ent_bound}
\end{equation}
where the meter state in Eq.~(\ref{ent_amp}) gives $I(\gamma)\approx4|\langle\hat{A}\rangle_\text{w}|^2$. The uncertainty bound clearly shows that the use of an $N$-partite entangled system enables the Heisenberg-limited precision scaling, achieving the precision improvement of $\sqrt{N}$ over the case of $N$ independent measurements in Eq.~(\ref{eq5}). 

\begin{figure}[t]
\centering
\includegraphics[width=1\columnwidth]{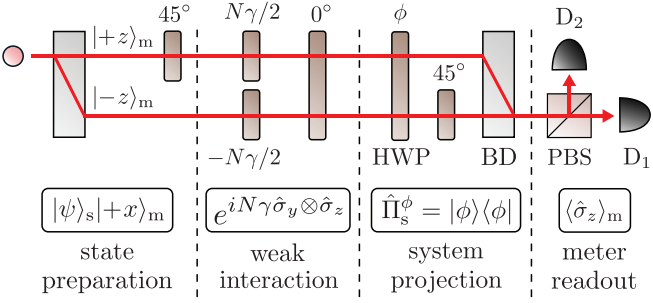}
\caption{Experimental schematic to estimate a path-dependent polarization change via WVA based on iterative interactions. The polarization mode and the path mode of a single photon are employed as the system and the meter, respectively, and they are initialized using the first beam displacer (BD) and half-wave plate (HWP). The path-dependent polarization change of $\gamma$ is implemented by rotating the next HWP in each path in opposite directions, and the $N$ iterative interactions are introduced by setting the HWP angles at $N \gamma/2$  and  $-N \gamma/2$, respectively. The system (polarization mode) is then projected onto $|\phi\rangle_\text{s}$ by post-selecting the photon coming from the lower path of the second BD, and the post-selection state is determined by the HWP angle $\phi$. The operation of the second BD converts the path mode state into the polarization mode state, so the meter readout is carried out using the polarizing beam splitter (PBS).}
\label{fig2} 
\end{figure}


We now present a WVA scheme which can achieve the Heisenberg-limit without using entangled resources. As shown in Fig.~\ref{fig1}(c), we consider $N$ iterative interactions, $\hat{U}^N=e^{iN\gamma\hat{A}\otimes\hat{\sigma}_z}$,  between the system and the meter prior to the detection. Interestingly, if the initial and post-selection system states are set to have the same relative amplitudes as those in Eq.~(\ref{ent_init}),
\begin{eqnarray}
\label{iterstate}
|\psi\rangle_\text{s}&=&\frac{1}{\sqrt{2}}(|\lambda_\text{max}\rangle_\text{s}+|\lambda_\text{min}\rangle_\text{s}),\\
|\phi\rangle_\text{s}&\approx&\frac{1}{\sqrt{2}|\langle\hat{A}\rangle_\text{w}|}\bigg[(\langle\hat{A}\rangle_\text{w}^*-N\lambda_{\text{min}})|\lambda_\text{max}\rangle_\text{s}\nonumber \\
&&\ \ \ \ \ \ \ \ \ \ \ \ \ \ \ \ -(\langle\hat{A}\rangle_\text{w}^*-N\lambda_\text{max})|\lambda_\text{min}\rangle_\text{s}\bigg], \nonumber
\end{eqnarray}
under the condition of Eq.~(\ref{condition}), the amplification factor is fixed to $\langle\hat{A}\rangle_\text{w}$ independently of $N$,
\begin{equation}
_\text{s}\langle\phi|\hat{U}^{N}|\psi\rangle_\text{s}|\!+\!x\rangle_\text{m}\approx\ \!_\text{s}\langle\phi|\psi\rangle_\text{s}(|\!+\!x\rangle_\text{m}+i\gamma\langle\hat{A}\rangle_\text{w}|\!-\!x\rangle_\text{m}),
\label{itermeter}
\end{equation}
and the system post-selection probability becomes identical to Eq.~(\ref{ent_prob}),
\begin{equation}
P_\text{iter}^{(N)}\approx |_\text{s}\langle\phi|\psi\rangle_\text{s}|^2\approx \frac{N^2(\lambda_\text{max}-\lambda_\text{min})^2}{4|\langle\hat{A}\rangle_\text{w}|^2}.
\label{iter_prob}
\end{equation}
Consequently, the quantum Cram\'{e}r-Rao bound for a single WVA measurement with $N$ iterative interactions also follows the Heisenberg-limited scaling, 
\begin{equation}
\Delta\gamma^{(N)}_\text{iter} \geq \frac{1}{\sqrt{P^{(N)}_\text{iter} I(\gamma)}}=\frac{1}{N|\lambda_\text{max}-\lambda_\text{min}|}.
\label{iterCramer}
\end{equation}
This result indicates that the perceived advantages of the entanglement-assisted WVA~\cite{Pang14,Pang15E,Chen19,Starek20} are in fact due to iterative local interactions between each particle of the entangled system and the meter, rather than coming from the entanglement itself.

\textit{Experiment}.---To demonstrate our WVA scheme based on iterative interactions, we experimentally estimate a small path-dependent polarization change. The experimental schematic is depicted in Fig.~\ref{fig2}. We employ the polarization mode and the path mode of a single-photon as the system and the meter, respectively. The system state can be written as a linear combination of the right circular $|\!+\!y\rangle_\text{s}$ and the left circular $|\!-\!y\rangle_\text{s}$ polarization states, and the upper and lower path states are denoted by $|\!+\!z\rangle_\text{m}$ and $|\!-\!z\rangle_\text{m}$, respectively. Here, $|\!\pm\! k\rangle$ have the eigenvalues of $\pm1$ for the Pauli-$k$ operator, $\hat{\sigma}_k$.  To initialize the system-meter state to, 
\begin{equation}
|\psi\rangle_\text{s}|\!+\!x\rangle_\text{m}=\frac{1}{\sqrt{2}}(|\!+\!y\rangle_\text{s}+|\!-\!y\rangle_\text{s})|\!+\!x\rangle_\text{m},
\end{equation}
we prepare a horizontally polarized single-photon in the path superposition, $|\!+\!x\rangle_\text{m}=(|\!+\!z\rangle_\text{m}+|\!-\!z\rangle_\text{m})/\sqrt{2}$, by using~a beam displacer (BD) and a half-wave plate (HWP). The details of the single-photon source are presented in Supplemental Material.

As shown in Fig.~\ref{fig2}, the weak interaction that causes a small path-dependent polarization change of $\gamma$ is implemented by rotating the angles of HWPs conditioned on the paths. The $N$ iterative interactions are introduced by setting the HWP angles at $N \gamma/2$ for  $|\!+\!z\rangle_\text{m}$ and  $-N \gamma/2$ for $|\!-\!z\rangle_\text{m}$. With an additional  HWP set at $0^\circ$, the conditional polarization operations are given as $e^{\pm iN\gamma\hat{\sigma}_y}$, and thus the iterative weak interactions are written as,
\begin{eqnarray}
\hat{U}^{N}&=&e^{iN\gamma \hat{\sigma}_y}\otimes{}_\text{m}|\!+\!z\rangle\langle+z|_\text{m}+e^{-i\gamma N\hat{\sigma}_y}\otimes{}_\text{m}|\!-\!z\rangle\langle-z|_\text{m} \nonumber\\
&=&e^{iN\gamma \hat{\sigma}_y\otimes\hat{\sigma}_z}.
\end{eqnarray}

For the amplified detection of $\gamma$, we project the system (the polarization mode) onto, 
\begin{equation}
|\phi\rangle_\text{s}=\frac{1}{\sqrt{2}}(|\!+\!y\rangle_\text{s}+e^{4i\phi}|\!-\!y\rangle_\text{s}),
\label{eq21}
\end{equation}
using the two HWPs at $\phi$ and $45^\circ$ and the second BD. Then, the meter (the path mode) state is found to be, 
\begin{equation}
_\text{s}\langle\phi|\hat{U}^N|\psi\rangle_\text{s}|\!+\!x\rangle_\text{m}\approx\ \!_\text{s}\langle\phi|\psi\rangle_\text{s}(|\!+\!x\rangle_\text{m}+i\gamma \langle N\hat{\sigma}_y\rangle_\text{w}|\!-\!x\rangle_\text{m}),
\label{eq22}
\end{equation}
where  the weak value of $\langle N \hat \sigma_y \rangle_\textrm{w}= {}_\text{s}\langle \phi | N\hat{\sigma}_y| \psi\rangle_\text{s} /{}_\text{s}\langle \phi | \psi\rangle_\text{s}$ can be large by choosing $\phi$ so that  $ {}_\text{s}\langle \phi | \psi\rangle_\text{s}  \approx 0$ at the limit of $|\gamma| \ll |1/\langle N\hat{\sigma}_y\rangle_\text{w}|$. The  interaction parameter $\gamma$ in the meter state can be obtained from the Pauli-$z$ expectation of the meter state:
\begin{equation}
\langle\hat{\sigma}_z\rangle_\text{m}\approx 2\gamma \ \! \text{Im}[\langle N\hat{\sigma}_y\rangle_\text{w}].
\label{Pauls}
\end{equation}
This shows that, in the meter readout, the interaction parameter $\gamma$ is amplified by the imaginary weak value $\text{Im}[\langle N\hat{\sigma}_y\rangle_\text{w}]$. As the operation of the second BD in Fig.~\ref{fig2} converts the path mode state into the polarization mode state, the meter readout is carried out using a polarizing beam splitter (PBS) and is estimated from $\langle\hat{\sigma}_z\rangle_\text{m}=(n_{1}-n_{2})/(n_1+n_2)$, where $n_1$ and $n_2$ are the photon counts detected at D$_1$ and D$_2$, repectively.

\begin{figure}[t]
\centering
\includegraphics[width=1\columnwidth]{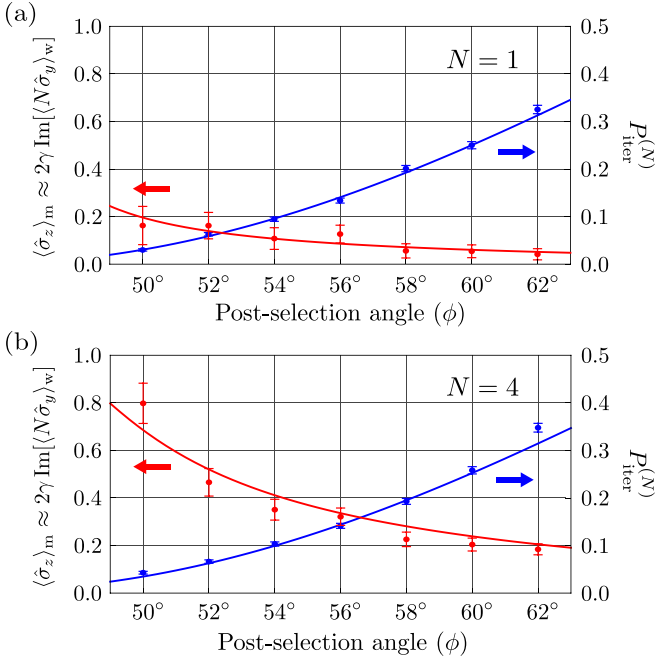}
\caption{The amplified meter readouts $\langle\hat{\sigma}_z\rangle_\text{m}\approx 2\gamma\ \!\text{Im}[\langle N\hat{\sigma}_y\rangle_\text{w}]$ and the detection probabilities $P^{(N)}_\text{iter}$ for the WVA measurements of $\gamma=1^{\circ}$ with (a) a single interaction and (b) four iterative interactions. As the post-selection state approaches the completely orthogonal state of the initial system state, $\phi=45^{\circ}$, the amplification effect on the meter expectation value $\langle\hat{\sigma}_z\rangle_\text{m}$ is enhanced while the detection probability decreases. The solid lines show the exact theoretical curves, and the error bars represent one standard deviation of Poissonian counting statistics.}
\label{fig3} 
\end{figure}

\begin{figure}[t]
\centering
\includegraphics[width=1\columnwidth]{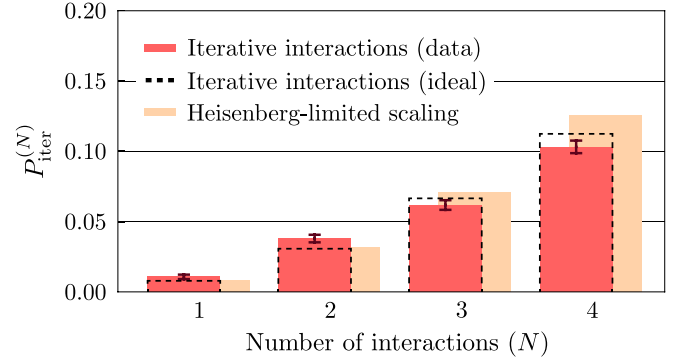}
\caption{The detection probabilities $P^{(N)}_\text{iter}$ of WVA based on $N$ iterative interactions. At each $N$, we adjust the post-selection state $|\phi\rangle_\text{s}$ according to Eq.~(\ref{iterstate}) in order to fix the amplification factor to $|\langle N\hat{\sigma}_y\rangle_\text{w}|=11.5$. For $N\!=\!1, 2, 3, 4$, the amplified meter expectation values for $\gamma=1^{\circ}$ are, respectively, measured as $\langle\hat{\sigma}_z\rangle_\text{m}=0.41\pm0.16$, $0.52\pm0.08$, $0.37\pm0.06$, $0.35\pm0.04$ with $\phi=47.5^{\circ}, 50.0^{\circ},52.3^{\circ}, 54.6^{\circ}$. The red bars represent the experimental data, and the dashed bars represent the ideally expected $P^{(N)}_\text{iter}$. The corresponding Heisenberg-limited scaling of $P^{(N)}_\text{H}\propto N^2$ is shown in the beige bars. The detection probabilities $P^{(N)}_\text{iter}$ have slight deviations from $P^{(N)}_\text{H}$ since the condition of Eq.~(\ref{condition}), $N\ll|\langle N\hat{\sigma}_y\rangle_\text{w}|=11.5$, does not hold as $N$ increases. This deviation can be elimilated by increasing the amplification factor. The error bars represent one standard deviation of Poissonian counting statistics.}
\label{fig4} 
\end{figure}

\textit{Experimental results}.---First, by varying the post-selection angle $\phi$ in Eq.~(\ref{eq21}), we study the trade-off relation between the meter expectation of $\langle\hat{\sigma}_z\rangle_\text{m}$ and the detection probability of $P^{(N)}_\text{iter}$ with $\gamma=1^\circ$ at $N=1$ and $4$. Figure~\ref{fig3} shows that the meter expectation is more amplified as the detection probability decreases. In addition, as expected from Eq.~(\ref{Pauls}), we observe that the WVA with $N=4$ iterative interactions provides nearly four times greater meter expectation values $\langle\hat{\sigma}_z\rangle_\text{m}$ compared to the WVA with $N=1$ while keeping the detection probability $P^{(N)}_\text{iter}$ almost independent of the number of interactions. The amplification factor is not exactly four times greater as the approximation condition, $|\gamma| \ll |1/\langle N\hat{\sigma}_y\rangle_\text{w}|$, does not satisfy well with a large amplification factor.

Next, we measure the enhanced detection probabilities from $N=1$  to 4 with a fixed amplification factor $|\langle N\hat{\sigma}_y\rangle_\text{w}|$. For the purpose of this demonstration, we set the amplification factor to 11.5 by varying the post-selection state $|\phi\rangle_\text{s}$ at each $N$ according to Eq.~(\ref{iterstate}). The experimental results in Fig.~\ref{fig4} show almost quadratic enhancement in the detection probability $P^{(N)}_\text{iter}$ with increased $N$. Given the same amplification factor in the meter state, as the minimum achievable uncertainty bound in Eq.~(\ref{iterCramer}) is inversely proportional to the square root of the detection probability $P^{(N)}_\text{iter}$, the quadratic enhancement of the detection probability results in the Heisenberg-limited precision scaling for $\Delta\gamma^{(N)}_\text{iter}$ by reducing statistical errors. Since the quadratic scaling of the detection probability is achieved under the condition in Eq.~(\ref{condition}), $N\ll|\langle N\hat{\sigma}_y\rangle_\text{w}|=11.5$, the experimental data in Fig.~\ref{fig4} have slight deviations from the Heisenberg-limited scaling as $N$ increases. However, this deviation can be eliminated by increasing the amplification factor.

\textit{Discussion}.---We have experimentally demonstrated that the Heisenberg-limited precision scaling can be achieved using a novel WVA scheme based on iterative interactions. This result indicates that the perceived advantages of the entanglement-assisted WVA~\cite{Pang14,Pang15E,Chen19,Starek20} are in fact due to iterative local interactions between each particle of an entangled system and a meter, rather than coming from the entanglement itself. Although quantum entanglement enables simultaneous interactions between all system particles and a meter, the fast interaction time is usually compromised by large decoherence which increases linearly with the size of the system~\cite{Dur04}. Therefore, the entanglement-assisted WVA does not offer a meaningful advantage over our scheme under decoherence~\cite{Boixo12,Demkowicz14}. Rather, since a large-scale entangled state is difficult to produce with high fidelity, our iterative interaction approach would provide practical applications in achieving the Heisenberg-limited precision scaling for weak-value amplification. 

\vspace{2mm}
This work was supported in part by the National Research Foundation of Korea (Grant No. 2019R1A2C3004812) and the ITRC support program (IITP-2021-2020-0-01606).



\begin{thebibliography}{}

\bibitem{Aharonov88} Y. Aharonov, D. Z. Albert, and L. Vaidman, Phys. Rev. Lett. \textbf{60}, 1351 (1988).

\bibitem{Dressel14} J. Dressel, M. Malik, F. M. Miatto, A. N. Jordan, and R. W. Boyd, Rev. Mod. Phys. \textbf{86}, 307 (2014).

\bibitem{Cho10} Y.-W. Cho, H.-T. Lim, Y.-S. Ra, and Y.-H. Kim, New J. Phys. \textbf{12}, 023036 (2010).


\bibitem{Hosten08} O. Hosten and P. Kwiat, Science \textbf{319}, 787 (2008).

\bibitem{Strubi13} G. Str\"ubi and C. Bruder, Phys. Rev. Lett. \textbf{110}, 083605 (2013).

\bibitem{Xu13} X.-Y. Xu, Y. Kedem, K. Sun, L. Vaidman, C.-F. Li, and G.-C. Guo, Phys. Rev. Lett. \textbf{111}, 033604 (2013).

\bibitem{Feizpour11} A. Feizpour, X. Xing, and A. M. Steinberg, Phys. Rev. Lett. \textbf{107}, 133603 (2011).

\bibitem{Hallaji17} M. Hallaji, A. Feizpour, G. Dmochowski, J. Sinclair, and A. M. Steinberg, Nat. Phys. \textbf{13}, 540 (2017).

\bibitem{Chen18} G. Chen \textit{et al.}, Nat. Commun. \textbf{9}, 93 (2018).


\bibitem{Combes14} J. Combes, C. Ferrie, Z. Jiang, and C. M. Caves, Phys. Rev. A \textbf{89}, 052117 (2014).

\bibitem{Ferrie14} C. Ferrie and J. Combes, Phys. Rev. Lett. \textbf{112}, 040406 (2014).

\bibitem{Knee14} G. C. Knee and E. M. Gauger, Phys. Rev. X \textbf{4}, 011032 (2014).

\bibitem{Zhang15} L. Zhang, A. Datta, and I. A. Walmsley, Phys. Rev. Lett. \textbf{114}, 210801 (2015).






\bibitem{Sinclair17} J. Sinclair, M. Hallaji, A. M. Steinberg, J. Tollaksen, and A. N. Jordan, Phys. Rev. A \textbf{96}, 052128 (2017).

\bibitem{Kedem12} Y. Kedem, Phys. Rev. A \textbf{85}, 060102(R) (2012).

\bibitem{Jordan14} A. N. Jordan, J. Mart\'{i}nez-Rinc\'{o}n, and J. C. Howell, Phys. Rev. X \textbf{4}, 011031 (2014).

\bibitem{Viza15} G. I. Viza, J. Mart\'{i}nez-Rinc\'{o}n, G. B. Alves, A. N. Jordan, and J. C. Howell, Phys. Rev. A \textbf{92}, 032127 (2015).




\bibitem{Harris17} J. Harris, R. W. Boyd, and J. S. Lundeen,  Phys. Rev. Lett. \textbf{118}, 070802 (2017).
\bibitem{Xu20} L. Xu, Z. Liu, A. Datta, G. C. Knee, J. S. Lundeen, Y.-Q. Lu, and L. Zhang,  Phys. Rev. Lett. \textbf{125}, 080501 (2020).



\bibitem{Pang15S} S. Pang and T. A. Brun, Phys. Rev. Lett. \textbf{115}, 120401 (2015).

\bibitem{Pang14} S. Pang, J. Dressel, and T. A. Brun, Phys. Rev. Lett. \textbf{113}, 030401 (2014).
\bibitem{Pang15E} S. Pang and T. A. Brun, Phys. Rev. A \textbf{92}, 012120 (2015).
\bibitem{Chen19} J.-S. Chen, B.-H. Liu, M.-J. Hu, X.-M. Hu, C.-F. Li, G.-C. Guo, and Y.-S. Zhang, Phys. Rev. A \textbf{99}, 032120 (2019).
\bibitem{Starek20} R. St\'arek, M. Mi\v{c}uda, R. Ho\v{s}\'{a}k, M. Je\v{z}ek, and J. Fiur\'{a}\v{s}ek, Opt. Express \textbf{28}, 34639 (2020).





\bibitem{Lundeen11} J. S. Lundeen, B. Sutherland, A. Patel, C. Stewart, and C. Bamber,  Nature \textbf{474}, 188 (2011).

\bibitem{Kim18} Y. Kim \textit{et al.},  Nat. Commun. \textbf{9}, 192 (2018).


\bibitem{Kocsis11} S. Kocsis \textit{et al.}, Science \textbf{332}, 1170 (2011).

\bibitem{Mahler16} D. H. Mahler \textit{et al.},  Sci. Adv. \textbf{2}, e1501466 (2016).

\bibitem{Cho19} Y.-W. Cho \textit{et al.},  Nat. Phys. \textbf{15}, 665 (2019).

\bibitem{Ramos20} R. Ramos, D. Spierings, I. Racicot, and A. M. Steinberg, Nature \textbf{583}, 529 (2020).

\bibitem{Kim21} Y. Kim \textit{et al.}, npj Quantum Inf. \textbf{7}, 13 (2021).

\bibitem{Kim12} Y.-S. Kim, J.-C. Lee, O. Kwon, and Y.-H. Kim, Nat. Phys. \textbf{8}, 117 (2012).


\bibitem{Braunstein94} S. L. Braunstein and C. M. Caves, Phys. Rev. Lett. \textbf{72}, 3439 (1994).

\bibitem{Braunstein96} S. L. Braunstein, C. M. Caves and G. J. Milburn, Ann. Phys. \textbf{247}, 135 (1996).











\bibitem{Giovannetti04} V. Giovannetti, S. Lloyd, and L. Maccone, Science \textbf{306}, 1330 (2004).








\bibitem{Dur04} W. D\"{u}r and H.-J. Briegel, Phys. Rev. Lett. \textbf{92}, 180403 (2004).

\bibitem{Boixo12} S. Boixo and C. Heunen, Phys. Rev. Lett. \textbf{108}, 120402 (2012).

\bibitem{Demkowicz14} R. Demkowicz-Dobrza\'{n}ski and L. Maccone, Phys. Rev. Lett. \textbf{113}, 250801 (2014).


\end{thebibliography}
\end{document}


\section*{\large{Supplemental Material: Heisenberg-limited metrology via weak-value amplification without using entangled resources}}

\vspace{10mm} 

\section{Quantum Fisher information}
The amount of information, that a pure quantum state $|M(\gamma)\rangle$ carries about an unknown parameter $\gamma$, can be estimated by the quantum Fisher information~\cite{Braunstein94,Braunstein96}, 
\begin{equation}
I(\gamma)=4\frac{d\langle M(\gamma)|}{d \gamma}\frac{d |M(\gamma)\rangle}{d \gamma}-4\left|\frac{d\langle M(\gamma)|}{d \gamma}|M(\gamma)\rangle\right|^2.
\end{equation}
 The quantum Fisher information of $|M(\gamma)\rangle\approx|\!+\!x\rangle+i\gamma\langle\hat{A}\rangle_\text{w}|\!-\!x\rangle$ is calculated to be,
 \begin{equation}
I(\gamma)\approx4|\langle\hat{A}\rangle_\text{w}|^2-4\gamma^2|\langle\hat{A}\rangle_\text{w}|^4\approx4|\langle\hat{A}\rangle_\text{w}|^2,
\end{equation}
 under the assumption of $|\gamma\langle\hat{A}\rangle_\text{w}|\ll1$.

\section{Single-photon source} A heralded single-photon source is utilized in the experiment, based on ultrafast-pumped spontaneous parametric down-conversion (SPDC) with a 1-mm type-II beamlike beta barium borate (BBO) crystal. The central wavelengths of the pump photon and the SPDC photon pair are 390 nm and 780 nm, respectively. The presence of a SPDC photon is heralded by detecting the other pair. The heralded single-photon source is collected via a single-mode fiber and transferred into the experimental setup shown in the main Fig. 2.

